\documentclass[12pt]{article}

\usepackage[body={17.5cm, 21cm},right=2cm]{geometry}

\usepackage{color}
\usepackage{graphicx}
\usepackage{epsf}
\usepackage{graphicx,epsfig}
\begin{document}
\vspace{1cm}
\begin{flushright}
CERN-PH-TH/2011-221
\end{flushright}

\vspace{5mm}
\vspace{0.5cm}
\begin{center}

\def\thefootnote{\fnsymbol{footnote}}

{\Large \bf Non-Gaussianity  in the Cosmic Microwave Background \\[0.3cm] Anisotropies at Recombination
in the Squeezed limit}
\\[1.5cm]
{\large  N. Bartolo$^{a,b}$, S. Matarrese$^{a,b}$ and A. Riotto$^{b,c}$}
\\[0.5cm]

\vspace{.3cm}
{\normalsize { \sl $^{a}$ Dipartimento di Fisica ``G.\ Galilei'', 
        Universit\`{a} di Padova, via Marzolo 8, I-131 Padova, Italy}}\\

\vspace{.3cm}
{\normalsize { \sl $^{b}$ INFN, Sezione di Padova, via Marzolo 8, I-35131 Padova, Italy}}\\

\vspace{.3cm}
{\normalsize {\sl $^{\rm c}$  CERN, Theory Division, CH-1211 Geneva 23, Switzerland}}\\

\vspace{.3cm}
{\normalsize {\sl E-mail: nicola.bartolo@pd.infn.it, sabino.matarrese@pd.infn.it and riotto@mail.cern.ch}}


\end{center}

\vspace{3cm}

\hrule \vspace{0.3cm}
{\small  \noindent \textbf{Abstract} \\[0.3cm]
\noindent We estimate analytically the second-order cosmic microwave background  temperature anisotropies at the recombination epoch in the squeezed limit and we deduce the  
contamination to the primordial local non-Gaussianity. We find that the level of contamination corresponds to $f_{\rm NL}^{\rm  con}={\cal O}(1)$ which is below the sensitivity of present experiments and smaller than the value ${\cal O}(5)$ recently claimed in the literature.

\vspace{0.5cm}  \hrule
\def\thefootnote{\arabic{footnote}}
\setcounter{footnote}{0}



\newpage 
\tableofcontents

\newcommand{\fix}{\Phi(\mathbf{x})}
\newcommand{\fiLx}{\Phi_{\rm L}(\mathbf{x})}
\newcommand{\fiNLx}{\Phi_{\rm NL}(\mathbf{x})}
\newcommand{\fik}{\Phi(\mathbf{k})}
\newcommand{\fiLk}{\Phi_{\rm L}(\mathbf{k})}
\newcommand{\fiLkone}{\Phi_{\rm L}(\mathbf{k_1})}
\newcommand{\fiLktwo}{\Phi_{\rm L}(\mathbf{k_2})}
\newcommand{\fiLkthree}{\Phi_{\rm L}(\mathbf{k_3})}
\newcommand{\fiLkfour}{\Phi_{\rm L}(\mathbf{k_4})}
\newcommand{\fiNLk}{\Phi_{\rm NL}(\mathbf{k})}
\newcommand{\fiNLkone}{\Phi_{\rm NL}(\mathbf{k_1})}
\newcommand{\fiNLktwo}{\Phi_{\rm NL}(\mathbf{k_2})}
\newcommand{\fiNLkthree}{\Phi_{\rm NL}(\mathbf{k_3})}

\newcommand{\kernel}{f_{\rm NL} (\mathbf{k_1},\mathbf{k_2},\mathbf{k_3})}
\newcommand{\dirac}{\delta^{(3)}\,(\mathbf{k_1+k_2-k})}
\newcommand{\dirackonektwokthree}{\delta^{(3)}\,(\mathbf{k_1+k_2+k_3})}

\newcommand{\beq}{\begin{equation}}
\newcommand{\eeq}{\end{equation}}
\newcommand{\beqarr}{\begin{eqnarray}}
\newcommand{\eeqarr}{\end{eqnarray}}

\newcommand{\angk}{\hat{k}}
\newcommand{\angn}{\hat{n}}

\newcommand{\tfnow}{\Delta_\ell(k,\eta_0)}
\newcommand{\tf}{\Delta_\ell(k)}
\newcommand{\tfone}{\Delta_{\el\ell_1}(k_1)}
\newcommand{\tftwo}{\Delta_{\el\ell_2}(k_2)}
\newcommand{\tfthree}{\Delta_{\el\ell_3}(k_3)}
\newcommand{\tffour}{\Delta_{\el\ell_1^\prime}(k)}
\newcommand{\deltatilde}{\tilde{\Delta}_{\el\ell_3}(k_3)}

\newcommand{\alm}{a_{\ell m}}
\newcommand{\almL}{a_{\ell m}^{\rm L}}
\newcommand{\almNL}{a_{\ell m}^{\rm NL}}
\newcommand{\almone}{a_{\el\ell_1 m_1}}
\newcommand{\almLone}{a_{\el\ell_1 m_1}^{\rm L}}
\newcommand{\almNLone}{a_{\el\ell_1 m_1}^{\rm NL}}
\newcommand{\almtwo}{a_{\el\ell_2 m_2}}
\newcommand{\almLtwo}{a_{\el\ell_2 m_2}^{\rm L}}
\newcommand{\almNLtwo}{a_{\el\ell_2 m_2}^{\rm NL}}
\newcommand{\almthree}{a_{\el\ell_3 m_3}}
\newcommand{\almLthree}{a_{\el\ell_3 m_3}^{\rm L}}
\newcommand{\almNLthree}{a_{\el\ell_3 m_3}^{\rm NL}}

\newcommand{\YLMstar}{Y_{L M}^*}
\newcommand{\Ylmstar}{Y_{\ell m}^*}
\newcommand{\Ylmstarone}{Y_{\el\ell_1 m_1}^*}
\newcommand{\Ylmstartwo}{Y_{\el\ell_2 m_2}^*}
\newcommand{\Ylmstarthree}{Y_{\el\ell_3 m_3}^*}
\newcommand{\Ylmstarfour}{Y_{\el\ell_1^\prime m_1^\prime}^*}
\newcommand{\Ylmstarfive}{Y_{\el\ell_2^\prime m_2^\prime}^*}
\newcommand{\Ylmstarsix}{Y_{\el\ell_3^\prime m_3^\prime}^*}

\newcommand{\YLM}{Y_{L M}}
\newcommand{\Ylm}{Y_{\ell m}}
\newcommand{\Ylmone}{Y_{\el\ell_1 m_1}}
\newcommand{\Ylmtwo}{Y_{\el\ell_2 m_2}}
\newcommand{\Ylmthree}{Y_{\el\ell_3 m_3}}
\newcommand{\Ylmfour}{Y_{\el\ell_1^\prime m_1^\prime}}
\newcommand{\Ylmfive}{Y_{\el\ell_2^\prime m_2^\prime}}
\newcommand{\Ylmsix}{Y_{\el\ell_3^\prime m_3^\prime}}

\newcommand{\comm}[1]{\textbf{\textcolor{rossos}{#1}}}
\newcommand{\lsim}{\,\raisebox{-.1ex}{$_{\textstyle <}\atop^{\textstyle\sim}$}\,}
\newcommand{\gsim}{\,\raisebox{-.3ex}{$_{\textstyle >}\atop^{\textstyle\sim}$}\,}

\newcommand{\jl}{j_\ell(k r)}
\newcommand{\jlfourone}{j_{\el\ell_1^\prime}(k_1 r)}
\newcommand{\jlfivetwo}{j_{\el\ell_2^\prime}(k_2 r)}
\newcommand{\jlsixthree}{j_{\el\ell_3^\prime}(k_3 r)}
\newcommand{\jlsix}{j_{\el\ell_3^\prime}(k r)}
\newcommand{\jlthree}{j_{\el\ell_3}(k_3 r)}
\newcommand{\jlthreetau}{j_{\el\ell_3}(k r)}

\newcommand{\Gaunt}{\mathcal{G}_{\el\ell_1^\prime \, \el\ell_2^\prime \, 
\el\ell_3^\prime}^{m_1^\prime m_2^\prime m_3^\prime}}
\newcommand{\Gaunttwo}{\mathcal{G}_{\el\ell_1^\prime \, \el\ell_2^\prime \, 
\el\ell_3}^{m_1^\prime m_2^\prime m_3}}
\newcommand{\Gauntstardef}{\mathcal{H}_{\el\ell_1 \, \el\ell_2 \, \el\ell_3}^{m_1 m_2 m_3}}
\newcommand{\Gauntstarone}{\mathcal{G}_{\el\ell_1 \, L \,\, \el\ell_1^\prime}
^{-m_1 M m_1^\prime}}
\newcommand{\Gauntstartwo}{\mathcal{G}_{\el\ell_2^\prime \, \el\ell_2 \, L}
^{-m_2^\prime m_2 M}}

\newcommand{\de}{{\rm d}}

\newcommand{\dangn}{d \angn}
\newcommand{\dangk}{d \angk}
\newcommand{\dangkone}{d \angk_1}
\newcommand{\dangktwo}{d \angk_2}
\newcommand{\dangkthree}{d \angk_3}
\newcommand{\dk}{d^3 k}
\newcommand{\dkone}{d^3 k_1}
\newcommand{\dktwo}{d^3 k_2}
\newcommand{\dkthree}{d^3 k_3}
\newcommand{\dkfour}{d^3 k_4}
\newcommand{\dallk}{\dkone \dktwo \dk}

\newcommand{\FT}{ \int  \! \frac{d^3k}{(2\pi)^3} 
e^{i\mathbf{k} \cdot \angn \eta_0}}
\newcommand{\planewave}{e^{i\mathbf{k \cdot x}}}
\newcommand{\dallkfourier}{\frac{\dkone}{(2\pi)^3}\frac{\dktwo}{(2\pi)^3}
\frac{\dkthree}{(2\pi)^3}}

\newcommand{\Bis}{B_{\el\ell_1 \el\ell_2 \el\ell_3}^{m_1 m_2 m_3}}
\newcommand{\Avbis}{B_{\el\ell_1 \el\ell_2 \el\ell_3}}

\newcommand{\los}{\mathcal{L}_{\el\ell_3 \el\ell_1 \el\ell_2}^{L \, 
\el\ell_1^\prime \el\ell_2^\prime}(r)}
\newcommand{\loszero}{\mathcal{L}_{\el\ell_3 \el\ell_1 \el\ell_2}^{0 \, 
\el\ell_1^\prime \el\ell_2^\prime}(r)}
\newcommand{\losone}{\mathcal{L}_{\el\ell_3 \el\ell_1 \el\ell_2}^{1 \, 
\el\ell_1^\prime \el\ell_2^\prime}(r)}
\newcommand{\lostwo}{\mathcal{L}_{\el\ell_3 \el\ell_1 \el\ell_2}^{2 \, 
\el\ell_1^\prime \el\ell_2^\prime}(r)}
\newcommand{\losfNL}{\mathcal{L}_{\el\ell_3 \el\ell_1 \el\ell_2}^{0 \, 
\el\ell_1 \el\ell_2}(r)}
\newcommand\ee{\end{equation}}

\newcommand\be{\begin{equation}}

\newcommand\eea{\end{eqnarray}}

\newcommand\bea{\begin{eqnarray}}
\def\d{d}
\def\C{{\rm CDM}}
\def\me{m_e}
\def\te{T_e}
\def\ti{\tau_{\rm initial}}
\def\tci#1{n_e(#1) \sigma_T a(#1)}
\def\tr{\eta_r}
\def\dtr{\delta\eta_r}
\def\dd{\tilde\Delta^{\rm Doppler}}
\def\dsw{\Delta^{\rm Sachs-Wolfe}}
\def\clsw{C_\ell^{\rm Sachs-Wolfe}}
\def\cldop{C_\ell^{\rm Doppler}}
\def\Dt{\tilde{\Delta}}
\def\mut{\mu}
\def\vt{\tilde v}
\def\hp{ {\bf \hat p}}
\def\sdv{S_{\delta v}}
\def\svv{S_{vv}}
\def\bvt{\tilde{\bv}}
\def\delt{\tilde{\delta_e}}
\def\cos{{\rm cos}}
\def\nn{\nonumber \\}
\def\bq{ {\bf q} }
\def\ba{ {\bf p} }
\def\bap{ {\bf p'} }
\def\bqp{ {\bf q'} }
\def\bp{ {\bf p} }
\def\bpp{ {\bf p'} }
\def\bk{ {\bf k} }
\def\bx{ {\bf x} }
\def\bv{ {\bf v} }
\def\qp{ p^{\mu}k_{\mu} }
\def\qpp { p^{\mu} k'_{\mu} }
\def\bgm{ {\bf \gamma} }
\def\bkp{ {\bf k'} }
\def\gq{ g(\bq)}
\def\gqp{ g(\bqp)}
\def\fp{ f(\bp)}
\def\h#1{ {\bf \hat #1}}
\def\fpp{ f(\bpp)}
\def\fz{f^{(0)}(p)}
\def\fpz{f^{(0)}(p')}
\def\f#1{f^{(#1)}(\bp)}
\def\fps#1{f^{(#1)}(\bpp)}
\def\dq{ {d^3\bq \over (2\pi)^32E(\bq)} }
\def\dqp{ {d^3\bqp \over (2\pi)^32E(\bqp)} }
\def\dpp{ {d^3\bpp \over (2\pi)^32E(\bpp)} }
\def\dtq{ {d^3\bq \over (2\pi)^3} }
\def\dtqp{ {d^3\bqp \over (2\pi)^3} }
\def\dtpp{ {d^3\bpp \over (2\pi)^3} }
\def\part#1;#2 {\partial#1 \over \partial#2}
\def\deriv#1;#2 {d#1 \over d#2}
\def\Done{\Delta^{(1)}}
\def\Dtwo{\tilde\Delta^{(2)}}
\def\fone{f^{(1)}}
\def\ftwo{f^{(2)}}
\def\tg{T_\gamma}
\def\delpp{\delta(p-p')}
\def\delb{\delta_B}
\def\tc{\eta_0}
\def\DD{\langle|\Delta(k,\mu,\eta_0)|^2\rangle}
\def\DDL{\langle|\Delta(k=l/\tc,\mu)|^2\rangle}
\def\bkpp{{\bf k''}}
\def\kmkp{|\bk-\bkp|}
\def\kmkpsq{k^2+k'^2-2kk'x}
\def\tt{ \left({\tau' \over \tau_c}\right)}
\def\kt{ k\mu \tau_c}

%
%
%

\section{Introduction}
Primordial non-Gaussianity (NG) of the cosmological perturbations has become a crucial aspect of both observational predictions of (inflationary) early universe models and of 
present and future observational probes of the Cosmic Microwave Background (CMB) anisotropies and Large-Scale Structures (LSS) of the universe. The main motivation is that 
detecting, or simply constraining, deviations from a Gaussian distribution of the primordial fluctuations allows to discriminate among different scenarios for the generation of the 
primordial perturbations.  A non-vanishing primordial NG encodes a wealth of information allowing to break  the degeneracy between models that, at the level of the power spectra, 
might result to be undistinguishable~\cite{ourreview}.  It is impressive that a future detection of a high level of primordial NG could rule out the standard single-field models of slow-roll 
inflation, which are characterized by weak gravitational interactions and thus predict tiny deviations from Gaussianity \cite{noi,Maldacena}.   
Statistics like the bispectrum (three-point correlation function) of CMB anisotropies~\cite{Lreview}, and higher-order statistics, and various observational probes of the LSS~\cite{CVM}, 
can be used to assess the level of primordial NG on various cosmological scales.  

It has now been fully appreciated the importance of distinguishing different {\rm shapes} of the primordial bispectrum, i.e. the dependence of the bispectrum on different triangular configurations 
formed by the three wave-vectors over which, in Fourier space, the primordial fluctuations are correlated~\cite{shape}. Different (classes of) inflationary models give rise to unique 
signals with specific triangular shapes, which thus probe different aspects of the physics of the primordial universe, in particular they can probe different kinds of interactions in the 
inflaton sector which are at the origin of the primordial non-Gaussian signal. For example, models in which non-linearities develop outside the horizon during inflation, or 
immediately after inflation, generate a level of NG which is local, as the NG part of the primordial curvature perturbation is a local function of the Gaussian part, being 
generated on superhorizon scales. This is generally the case when a light scalar field, different from the inflaton, contributes to the curvature perturbation (for example, 
multi-field models of inflation~\cite{two}, curvaton~\cite{luw,ngcurv} and inhomogenoeus reheating models~\cite{hk,varcoupling,varmass}). 
In momentum space, the three point function, or bispectrum, arising from the local NG is dominated by the so-called ÒsqueezedÓ configuration, where one of the momenta is
much smaller than the other two ($k_1 \ll k_2 \sim k_3$) and it is parametrized by the non-linearity parameter $f^{\rm loc}_{\rm NL}$. ``Equilateral'' NG, peaking for $k_1 \sim  k_2 \sim k_3$ 
is generally predicted by inflationary models  where the inflaton field is characterized by derivative interactions so that the correlation between the modes peaks 
when all the modes cross the horizon during inflation at the same time (see, e.g., inflationary models with non-standard kinetic term ~\cite{Chengeneral}, or DBI 
models of inflation~\cite{DBI}). For other interesting NG shapes/models (such as ``orthogonal'' ~\cite{orthogonal} and ``folded'' (or flattened) shapes~\cite{Chengeneral,H,Fasiello,Corasaniti,Cnewshapes}) 
we refer the reader to the review~\cite{Chenreview}. Present limits on NG are summarized by the WMAP measurements $- 10 < f^{\rm loc}_{\rm NL} < 74$  and $- 214 < f^{\rm equi}_{\rm NL} < 266$ at $95 \%$  CL~\cite{WMAP7} (see~\cite{FLS} for CMB constraints on other types of NG).  The \emph{Planck} satellite~\cite{planck} is expected to improve the sensitivity to the non-linearity parameters by an order 
of magnitude. In particular, for the local case, it is expected to be sensitive to a value of the non-linearity parameter as low as $f^{\rm loc}_{\rm NL}=3$ at 1$\sigma$ (accounting also for polarization information; for the equilateral case 
$f^{\rm equil}_{\rm NL}=30$)~\cite{ks,BZ,LY,SL,Baumann}.  

However there are many sources of NG in the CMB anisotropies beyond the primordial one, including systematic effects and astrophysical contamination. 
Given the deep implications that a detection of primordial non-Gaussianity would have, it is crucial that all  possible sources of contamination for the 
primordial signal are well understood and under control. In fact any non-linearities can make initially Gaussian perturbations non-Gaussian. 
Such non-primordial effects can thus complicate the extraction of the primordial non-Gaussianity. 
One of their main effects could be that of ``mimicking'' a three-point correlation function that is similar in shape to the primordial one, 
thus inducing a bias to the estimator of primordial NG used.  Therefore we have to be sure we are not ascribing a primordial origin to a signal that is extracted from the CMB (or LSS) data
using estimators of non-Gaussianity when that signal has a different origin. Moreover we must always specify which primordial non-Gaussianity we study the contamination
from the non-primordial sources (e.g. primordial non-Gaussianity of ``local'', ``equilateral'' or ``folded'' shape).
In that respect the reasonable question to ask is which kind of primordial non-Gaussianity a given secondary 
effect can contaminate most, wether it is of the so-called Òlocal typeÓ
or of the Òequilateral typeÓ. 

As far as CMB anisotropies are concerned, well-known examples of contamination are the NG produced by (non-linear) secondary anisotropies, like, e.g., the Sunyaev-ZelÕdovich effect,
gravitational lensing, Integrated-Sachs-Wolfe (or Rees-Sciama)~(see, e.g.~\cite{Aghanim} for a review on CMB secondary anisotropies). 
Up to now it has been found that the most serious contamination to the local type of primordial NG is the 
the coupling between the Integrated Sachs-Wolfe/Rees-Sciama effect and the weak gravitational lensing \cite{GoldSperg}.
In the detailed analysis of~\cite{Hanson,MangilliVerde,LCH}  it is shown that the ISW/Rees-Sciama-lensing bispectrum produces a bias of $f_{\rm NL}^{\rm cont} \simeq 10$ 
to local primordial non-Gaussianity, while the bias to equilateral primordial non-Gaussianity is negligible (see also~\cite{SerraCooray,SZ}). 
Other secondary effects have a little impact on the (local) primordial NG with $|f_{\rm NL}| \leq 0.7$  (for a summary of the level of contamination 
see the reviews~\cite{review,Lreview} and references therein).

Another relevant source of NG comes from the physics operating at the recombination epoch.
The interest in these sources of contamination is due to the fact that they actually bring previously unknown effects coming from the non-linearities in the Boltzmann 
equations. A naive estimate would tell that these non-linearities are tiny being suppressed by an extra power of the gravitational
potential. However, the dynamics at recombination is quite involved because all the non-linearities in the
evolution of the baryon-photon fluid at recombination and the ones coming from general relativity should
be accounted for. This complicated dynamics might lead to unexpected suppressions or enhancements of
the NG at recombination. A step towards the evaluation of the three-point correlation function has been
taken in Ref.~\cite{Crec} where some effects were taken into account in the squeezed triangle limit,
corresponding to the case when one wavenumber is much smaller than the other two and was outside
the horizon at recombination. Refs.~\cite{CMB2I,CMB2II,P,STZI,Beneke}, present the computation of the full system
of Boltzmann equations, describing the evolution of the photon, baryon and Cold Dark Matter (CDM)
fluids, up to second order  in the perturbations. These equations allow to follow the time evolution of
the CMB anisotropies at second order on all angular scales from the early epochs, when the cosmological
perturbations were generated, to the present time, through the recombination era. These calculations
set the stage for the computation of the full second-order radiation transfer function at all scales and
for a a generic set of initial conditions specifying the level of primordial non-Gaussianity \footnote{Of course,
for specific effects on small angular scales like Sunyaev-ZelÕdovich, gravitational lensing, etc., fully nonlinear
calculations would provide a more accurate estimate of the resulting CMB anisotropy, however, as
long as the leading contribution to second-order statistics like the bispectrum is concerned, second-order
perturbation theory suffices.}.

There has been some debate about the level of contamination that the non-linear evolution at recombination can generate. A first work, partially implementing numerically the 
second-order Boltzmann equations in the small scale limit, claimed a very high level of contamination to the local-type $f_{\rm NL}$~\cite{frI}. However it has been shown in 
Ref.~\cite{BRRec} that the source of this effect, the small-scale non-linear evolution of the second-order gravitational potential at recombination, actually contaminates mainly the 
{\rm equilateral}-type NG. A contamination of $f_{\rm NL}^{\rm cont}(\rm equil) \simeq 5$ for an experiment like \emph{Planck} was found~\cite{BRRec}, which is actually low 
w.r.t. the expected \emph{Planck} sensitivity to this kind of NG. The contamination to the local NG from the small scale evolution of the second-order gravitational potential is also very low, 
 $f_{\rm NL}^{\rm cont}(\rm loc) \simeq 0.3$~\cite{BRRec} for an experiment like \emph{Planck}. 
In Ref.~\cite{Nitta} another numerical implementation of the second-order Boltzmann equations has been performed to compute 
the contamination from specific terms appearing in the 
second-order radiation transfer function. These are the terms that can be written (in the Poisson gauge adopted in Ref.~\cite{Nitta}) as products of the first-order perturbations 
(in the form, e.g., $(\Phi^{(1)})^2$, where $\Phi^{(1)} $ is the linear gravitational potential). The contamination to local NG was found to be negligible for an experiment like \emph{Planck} 
being $|f_{\rm NL}^{\rm cont}(\rm loc) | <1$~\cite{Nitta}. 

Recently, Ref.~\cite{frII} analysed numerically the second-order evolution of the photon-baryon system at recombination claiming an overall contamination 
$f_{\rm NL}^{\rm cont}(\rm loc) = 5$ up to a maximum multipole $\ell_{\rm max}=2000$.  
Such a contamination must therefore come from terms in the second-order radiaton transfer function that are ``intrinsically'' of second order. 
The authors of Ref.~\cite{frII} suggest that the main candidate is the second-order monopole perturbation  $(\Theta^{(2)}_{\rm SW}=\Delta^{(2)}_{00}/4+\Phi^{(2)})$, 
which is the usual term appearing in the CMB anisotropies due to the intrinsic photon 
energy density fluctuations  $\Delta^{(2)}_{00}$ and the gravitational redshift due to the potential, both  evaluated at recombination.
In fact such a term on large scales reduces to the Sachs-Wolfe effect. There has been some debate 
on how such a second-order contribution is able to produce a contamination $f_{\rm NL}^{\rm cont}(\rm loc) = 5$ (see, e.g., \cite{KHunting,Benasque}). 
For example it is not clear what is the physical mechanism generating such a squeezed signal: if it occurs due to an integration over the terms (squared in first-order perturbations) that  source the ``intrinsically'' second-order
monopole, then, naively, one does not expect that a squeezed signal can be generated, because such a contribution would be associated to a causal evolution on small scales.  

The goal of this paper is to present a transparent computation of the bispectrum in the squeezed limit through a convenient coordinate rescaling~\cite{Crec,beyond,B}. To understand such a rescaling, it is important to recall what is generally the origin of a squeezed non-Gaussian signal: typically the local-form bispectrum is generated when  short-wavelength fluctuations are modulated by long-wavelength fluctuations. In particular we will focus on the temperature anisotropies at recombination when the long wavelength mode is outside the horizon, but observable at the present epoch.  Thus, the effect of the long wavelength mode imprinted at recombination can be described simply by a coordinate transformation. In this way we can describe in a simple way the coupling of small scales to large scales that can generally produce the local form bispectrum. A similar cross-talk between large and small scales gives rise to the ISW-lensing cross-correlation bispectrum. 

The plan of the paper is the following: In Sec. II we show that a simple coordinate rescaling reproduces correctly the second-order expressions for the temperature anisotropies in the squeezed limit. In Sec. III we take advantage of this coordinate rescaling to estimate the bispectrum from the CMB anisotropies at recombination in the squeezed limit and the corresponding contamination to the local primordial non-Gaussianity. Sec. IV contains our conclusions.


\section{Second-order CMB anisotropy at recombination}
\noindent
To motivate the fact that the rescaling of coordinates (that we will soon introduce) correctly captures the physics of the CMB anisotropies at second order in the squeezed limit, we will first  
show that such a rescaling applied to the linear equations properly reproduces the second-order sources found in Refs.~\cite{review,CMB2I,CMB2II}. 
We first recall the Boltzmann 
equations at first-order in perturbation theory for the CMB anisotropies. Since this subject
is rather standard, we refer the reader to standard books for more details~\cite{Dodelsonbook} (see also Refs.~\cite{review,CMB2II}).
Our starting metric is

\begin{equation}
\de s^2=a^2(\eta)\left[-e^{2\Phi}\de \eta^2+e^{-2\Psi}\de{\bf x}^2\right]\, ,
\end{equation}
where $a(\eta)$ is the scale factor as a function of the conformal time $\eta$, and we have neglected vector and tensor perturbations. 
The equations of motion of the first two moments of the Boltzmann equations for the CMB photons are 
\begin{equation}
\label{LH2B1l}
\Delta^{(1)'}_{00}+\frac{4}{3} \partial_i 
v^{(1)i}_\gamma-4\Psi^{(1)'}=0\, ,
\end{equation}
\begin{equation}
\label{LH2B2l}
v^{(1)i\prime}_\gamma+\frac{3}{4} \partial_j 
\Pi^{(1)ji}_\gamma+\frac{1}{4} \Delta^{(1),i}_{00}
+\Phi^{(1),i}=-\tau' \left( v^{(1)i}-v^{(1)}_\gamma \right)\, , 
\end{equation}
where $\Pi^{ij}$ is the photon quadrupole moment and $\tau'=-\overline{n}_e\sigma_T a$ is the
differential optical depth in terms of the average number of electron number density $\overline{n}_e$ and Thomson cross section $\sigma_T$. 
Here the primes indicate differentiation with respect to $\eta$ and $\partial_i$ differentiation w.r.t. $x^i$.
The two equations above are complemented by the momentum 
continuity equation for baryons, which can be conveniently written 
as  
\begin{equation}
\label{LH2bv1}
v^{(1)i}=v^{(1)i}_\gamma+\frac{R}{\tau'} \left[v^{(1)i\prime}+
{\cal H} v^{(1)i} +\Phi^{(1),i}   \right]\, ,
\end{equation}
where we have introduced the baryon-photon ratio 
$R \equiv 3 \rho_b/(4\rho_\gamma)$ and we have indicated by ${\cal H} =a'/a$ the Hubble rate in conformal time. 

Eq.~(\ref{LH2bv1}) is in a form ready for a consistent expansion 
in the small quantity $\tau^{-1}$ which can be performed 
in the tight-coupling limit. By first taking $v^{(1)i}=v^{(1)i}_\gamma$ 
at zero order and then using this relation in the 
left-hand side of Eq.~(\ref{LH2bv1}) one obtains
\begin{equation}
\label{LH2vv}
v^{(1)i}-v^{(1)i}_\gamma=\frac{R}{\tau'} \left[v^{(1)i\prime}_\gamma
+{\cal H} v^{(1)i}_\gamma +\Phi^{(1),i}   \right]\, .
\end{equation}
Such an expression for the difference of velocities can be used in 
Eq.~(\ref{LH2B2l}) to give the evolution equation for 
the photon velocity in the limit of tight coupling
\begin{equation}
\label{LH2vphotontight}
v^{(1)i\prime}_\gamma+{\cal H}\frac{R}{1+R}v^{(1)i}_\gamma 
+\frac{1}{4} \frac{\Delta^{(1),i}_{00}}{1+R}+\Phi^{(1),i} =0\, .
\end{equation}
Notice that in Eq.~(\ref{LH2vphotontight}) 
we are neglecting the quadrupole of the photon distribution $\Pi^{(1) ij}$ 
(and all the higher moments) since it is well known that at linear 
order such moment(s) are suppressed in the tight-coupling limit 
by (successive powers of) $1/\tau'$ with respect to the first two 
moments, the photon energy density and velocity. 
Eqs.~(\ref{LH2B1l}) and (\ref{LH2vphotontight}) are the master equations 
which govern the photon-baryon fluid acoustic 
oscillations before the epoch of recombination when photons and baryons 
are tightly coupled by Compton scattering.

In the tight-coupling limit the contribution to the CMB anisotropies from recombination is given by
\be
\label{tot}
\Theta^{(1)}({\bf k},{\bf n},\eta_0)=\left(\frac{1}{4} \Delta^{(1)}_{00}+ \Phi^{(1)}+{\bf v}^{(1)} \cdot {\bf n} \right)_{\eta=\eta_{\rm rec}}\, .
\ee
We are now interested in computing the second-order CMB anisotropies in the squeezed limit, i.e. when one of the momenta is much smaller than the other two, $k_1\ll k_2\simeq k_3$. 
In particular, we will be interested in the case in which the smallest momentum is at most  equal to the wavenumber entering the horizon at equivalence, $k_1\lsim k_{\rm eq}$. 

Instead of solving the complicated network of second-order Boltzmann equations for the CMB temperature anisotropies, we use the following trick. As the wavenumber $k_1\lsim k_{\rm eq}$ 
corresponds to a perturbation which is almost larger than the horizon at recombination and the evolution in time of the corresponding gravitational potential is very moderate 
(one can easily check, for instance,  that $\Phi^{(1)}_{{\bf k}_1}(\eta)$ changes its magnitude by at most 10\% during the radiation epoch for $k_1=k_{\rm eq}$), we can absorb the large-scale perturbation
with wavelength $\sim k_1^{-1}$ in the metric by redefining the time and the space coordinates as follows. Let us indicate with $\Phi_\ell$ and $\Psi_\ell$ the parts of the 
gravitational potentials that receive contributions only from the large-scale modes $k_1 \lsim k_{\rm eq}$ (see, e.g., \cite{beyond}). If the scale factor is a power law 
$a(\eta)\propto\eta^\alpha$ ($\alpha=1$ and $\alpha=2$ for the period of radiation and matter domination, respectively), we can perform the redefinitions

\begin{equation}
\label{rescaling1}
a^2(\eta)e^{2\Phi_\ell}\de \eta^2=\eta^{2\alpha}e^{2\Phi_\ell}\de\eta^2=\overline{\eta}^{2\alpha}\de\overline{\eta}^{2}=a^2(\overline{\eta})\de\overline{\eta}^2\Rightarrow \overline{\eta}=e^{\frac{1}{1+\alpha}\Phi_\ell}\eta\, ,
\end{equation}
and

\be
\label{rescaling2}
a^2(\eta)e^{-2\Psi_\ell}\de {\bf x}^2=\eta^{2\alpha}e^{-2\Psi_\ell}\de {\bf x}^2=
\overline{\eta}^{2\alpha}e^{\frac{-2\alpha}{1+\alpha}\Phi_\ell}e^{-2\Psi_\ell}\de {\bf x}^2
=a^2(\overline{\eta})\de\overline{{\bf x}}^2\Rightarrow \overline{{\bf x}}=e^{\frac{-\alpha}{1+\alpha}\Phi_\ell}e^{-\Psi_\ell}{\bf x}\, .
\ee
In particular, the combination 
\be
\overline{k}\overline{\eta}=e^{\Phi_\ell+\Psi_\ell}\, k\eta\, ,
\ee
where $\overline{k}$ and $k$ are the wavenumbers in the two coordinate systems. Obviously, if one wishes to account for the fact that at recombination the universe is not fully matter-dominated, 
one should perform a more involved coordinate transformation which will eventually depend also on the parameter $R$.
To test the goodness of this procedure, let us consider for instance the equation of motion of the gravitational potential $\Psi_{\bf k}(\eta)$ at second-order during the radiation epoch. 
From the full second-order equations (B.13) and (B.14) of Ref. \cite{review}, we see that this equation reads in the squeezed limit 
\begin{equation}
\label{b}
\Psi^{(2)''}_{{\bf k}}+4{\cal H}\Psi^{(2)'}_{{\bf k}}+c_s^2k^2\Psi^{(2)}_{{\bf k}}=-4c_s^2\left(\Psi^{(1)}_{{\bf k}_1}+\Phi^{(1)}_{{\bf k}_1}\right)k_2^2\Psi^{(1)}_{{\bf k}_2}\, ,
\end{equation}
where in  the source in the r.h.s. we have implicitly assumed
an integration over the momenta ${\bf k}_1$ and ${\bf k}_2$ with the corresponding Dirac delta to ensure momentum conservation. We will adopt this convention from now on. In fact, for a 
bispectrum computation, it is the kernel of the source term that  matters, and from this point of view we can identify the wavenumber ${\bf k}_1$ as a long wavelength modulation of the perturbation 
in the momentum ${\bf k}_2$, so that  ${\bf k}={\bf k}_1+{\bf k}_2\simeq {\bf k}_2$.  

It is reassuring that this equation can be simply obtained by rescaling the coordinates, following  (\ref{rescaling1}) and (\ref{rescaling2}).  Just consider the (linear) equation of 
motion~\cite{Dodelsonbook,CMB2II}
\begin{equation}
\label{Psirad}
\Psi^{(1)''}+4{\mathcal H}\Psi^{(1)'}-c_s^2\nabla^2\Psi^{(1)}=0\, ,
\end{equation}
whose solution is 
\begin{equation}
\label{psirad}
\Psi_{\bf k}^{(1)}(\eta)=3\Psi_{\bf k}^{(1)}(0)\frac{\left(-c_s k \eta\,\cos(c_s k \eta)+\sin(c_s k \eta)\right)}{(c_s k \eta)^3}\, .
\end{equation}
Applying  (\ref{rescaling1}) and (\ref{rescaling2}) to Eq.~(\ref{Psirad}) we obtain 
\begin{equation}
\label{P2radeq}
\Psi^{''}+4{\mathcal H}\Psi^{'}- c_s^2 e^{2(\Phi_\ell+\Psi_\ell)} \nabla^2 \Psi=0\,. 
\end{equation}
Now, expanding the perturbations into a first- and second-order parts, $\Psi=\Psi^{(1)}+\Psi^{(2)}/2$, and identifying in Fourier space the long wavelength mode with  ${\bf k}_1$, one can easily check that Eq.~(\ref{P2radeq}) exactly correponds to Eq.~(\ref{b}).  The solution of Eq. (\ref{b}) is
\begin{eqnarray}
\label{c}
\Psi^{(2)}_{{\bf k}}(\eta)&=&3\Psi_{\bf k}^{(2)}(0)\frac{\left(-c_s k \eta\,\cos(c_s k \eta)+\sin(c_s k \eta)\right)}{(c_s k \eta)^3}\nonumber\\
&+&6\,\frac{3 \,c_s k \eta\,\cos(c_s k \eta)+(-3+(c_s k \eta)^2)\,\sin(c_s k \eta)}{(c_s k \eta)^3}\,\Psi^{(1)}_{{\bf k}_1}(0)\Psi^{(1)}_{{\bf k}_2}(0)\, .
\end{eqnarray}
Again, to have a further check of our procedure, we point out that this solution can be found by employing the coordinate rescalings  (\ref{rescaling1}) and (\ref{rescaling2}). Taking the expression (\ref{psirad}) and assuming that
$\Psi^{(1)}_{{\bf k}}\simeq \Phi^{(1)}_{{\bf k}}$ we find

\begin{eqnarray}
\label{phiradrescaled}
\Psi_{\bf k}(\eta)&=&3\Psi_{ \overline{{\bf k}}}(0)\frac{\left(-c_s \overline{k}_2 \overline{\eta}\,\cos(c_s \overline{k}_2\overline{ \eta})+\sin(c_s \overline{k}_2 \overline{\eta})\right)}{(c_s \overline{k}_2 \overline{\eta})^3}\nonumber\\
&=&3\Psi_{\overline{{\bf k}}}(0)\frac{\left(-e^{2{\Psi^{(1)}_{{\bf k}_1}}} c_s k_2 \eta\,\cos\left(e^{2{\Psi^{(1)}_{{\bf k}_1}}}c_s k_2\eta\right)+\sin\left(e^{2{\Psi^{(1)}_{{\bf k}_1}}} c_s k_2 \eta\right)\right)}{\left(e^{2{\Psi^{(1)}_{{\bf k}_1}}} c_s k_2 \eta\right)^3}\, .
\end{eqnarray}
Expanding at second-order one, one recovers exactly the expression (\ref{c})\footnote{Notice that here we are using the short-hand notation 
$\overline{k}\overline{\eta}=e^{\Phi^{(1)}_{{\bf k}_1}+\Psi^{(1)}_{{\bf k}_1}}\, k\eta$ which simplifies the computation giving nonetheless the correct result. Notice that the coordinate rescaling 
also transforms the gravitational potential $\Psi_{{\bf k}}(0)$. This change however is reabsorbed in the physical initial conditions.}. 

We have also verified that using this procedure applied to Eqs.~(\ref{LH2B1l}) and~(\ref{LH2vphotontight}) one reobtains  the corresponding second-order Boltzmann equations already computed in Eqs.~(205) and~(206) of Ref.~\cite{review} expressed in terms of the bolometric temperature, i.e., in terms of the variable $[\Delta^{(2)}-(\Delta^{(1)})^2]$. The bolometric temperature is defined as (see, e.g.,~\cite{frII,Nitta}) $(T/\bar{T})^4\equiv \Delta $, where $\Delta({\bf x}, {\bf n}, \tau)=\int dp p^3 f/(\int dp p^3 \bar{f})$ is the (normalized) brightness function, $f$ being the photon distribution function. By Taylor expanding this definition one finds 
\begin{equation}
\Delta^{(2)}=4 \Theta^{(2)}+16 \left( \Theta^{(1)} \right)^2=4 \Theta^{(2)}+\left( \Delta^{(1)} \right)^2\, ,
\end{equation}
where, in the notations of Refs.~\cite{Nitta,review}, 

\begin{equation}
\label{bol}
T=\overline{T}e^\Theta=T\, ,\,\,\,\,\frac{\Delta T}{\overline{T}}=\Theta^{(1)}+\frac{1}{2}\Theta^{(2)}+\frac{1}{2}\left(\Theta^{(1)}\right)^2\, ,
\end{equation}
is the bolometric temperature so that one recovers the temperature variables used in~\cite{Nitta,review} 
(notice that, despite using the same symbol $\Theta$ this temperature is different from the brightness temperature defined, e.g., in Eq.~(4.1) of~\cite{frII}). Since the existing literature usually computed the CMB bispectrum in terms of the bolometric temperature (e.g., \cite{Nitta,frII}), our results for the CMB bispectrum will refer to this variable. Let us show in some details how the coordinate rescaling method works for Eqs.~(\ref{LH2B1l}) and~(\ref{LH2vphotontight}). Under the coordinate rescalings  (\ref{rescaling1}) and (\ref{rescaling2})
Eq.~(\ref{LH2B1l}) transforms as
 \be 
 4 \Theta'_{00}+\frac{4}{3} e^{\Phi+\Psi} \partial_i v_{\gamma}^i-4 \Psi'=0\, .
 \ee
Expanding this equation at second-order in the perturbations (and expressing the photon velocity in terms of $\Delta_{00}$ and $\Psi$) one finds
\be 
\label{D2eqsqueezed}
4 {\Theta^{(2)}_{00}}'+\frac{4}{3} \partial_i v^{i(2)}_{\gamma}-4 \Psi^{'(2)}=- 2 (\Phi^{(1)}+\Psi^{(1)}) (4\Psi^{(1)'}-4{\Theta^{(1)}_{00}}' ) \, .
\ee
This equation can be confronted with Eq.~(205) of Ref.~\cite{review}
\begin{equation}
\label{D2eq}
\Delta^{(2)'}_{00}+\frac{4}{3} \partial_i v^{(2)i}_\gamma-4\Psi^{(2)'}={\cal S}_\Delta\, ,
\end{equation}
where the source term is given by
\begin{eqnarray}
\label{SD}
{\cal S}_\Delta&=&\left( \Delta^{(1)2}_{00} \right)^{\prime}-2(\Phi^{(1)}+\Psi^{(1)})(4\Psi^{(1)\prime}
-\Delta^{(1)\prime}_{00})-\frac{8}{3} v^{(1)i}_\gamma (\Delta^{(1)}_{00}+4\Phi^{(1)})_{,i}
\nonumber \\
&+&\frac{16}{3}(\Phi^{(1)}+\Psi^{(1)})^{,i}v_i-\frac{8}{3}R 
\left( \frac{\cal H}{1+R} v^{(1)2}_\gamma-\frac{1}{4}\frac{v^{(1)}_{\gamma i} \Delta^{(1),i}_{00}}{1+R}\right)
\, . \nonumber \\
\end{eqnarray}
It is easy to check that in the squeezed limit, $k_1 \ll k_2 \simeq k_3$, Eq.~(\ref{D2eq}) exactly coincides to Eq.~(\ref{D2eqsqueezed}) when written in terms of the bolometric monopole
$[\Delta^{(2)}_{00}-(\Delta^{(1)}_{00})^2]$. Similarly one can check that the squeezed limit of Boltzmann equation for the second-order velocity equation of the photon-baryon system can be obtained in the same way. Let us start from the (linear) Eq.~(\ref{LH2vphotontight}). By applying the coordinate rescalings  (\ref{rescaling1}) and (\ref{rescaling2}) we obtain
\be
\label{vtightsqueezed}
v^{\prime}_\gamma+{\cal H}\frac{R}{1+R}v^{i}_\gamma 
+\frac{1}{4} e^{(\Phi+\Psi)}\frac{4 \Theta^{,i}_{00}}{1+R}+ e^{(\Phi+\Psi)}\ \Phi^{(1),i} =0\, ,
\ee
which, expanded at second-order in the perturbations, becomes 
\begin{eqnarray}
\label{vtightsqueezed2nd}
v^{(2)i\prime}_\gamma +{\cal H} \frac{R}{1+R}v^{(2)i}_\gamma
+\frac{1}{4} \frac{4 \Theta^{(2),i}_{00}}{1+R}+\Phi^{(2),i}=-\frac{1}{2(1+R)} (\Phi^{(1)}+\Psi^{(1)}) 4 \Theta^{,i}_{00}\ - 2 (\Phi^{(1)}+\Psi^{(1)}) \Phi^{(1),i}\, .
\end{eqnarray}
This equation corresponds to the squeezed limit of the velocity continuity equation computed directly at second order in Ref.~\cite{review} (see Eq.~(209))
\begin{eqnarray}
\label{v2eqf}
v^{(2)i\prime}_\gamma +{\cal H} \frac{R}{1+R}v^{(2)i}_\gamma
+\frac{1}{4} \frac{\Delta^{(2),i}_{00}}{1+R}+\Phi^{(2),i}={\cal S}^{i}_V\, ,
\end{eqnarray}
where
\begin{eqnarray}
\label{SV}
{\cal S}^{i}_V&=&-\frac{3}{4(1+R)} \partial_j \Pi^{(2)ji}_\gamma -2 \omega'_i-2{\cal H} \frac{R}{1+R} \omega^i
+2 \frac{{\cal H}R}{(1+R)^2} \Delta^{(1)}_{00} v^{(1)i}_\gamma
\nonumber \\
&+&\frac{1}{4(1+R)^2}\left(  \Delta^{(1)2}_{00} \right)^{,i}+\frac{8}{3(1+R)}v^{(1)i}_\gamma \partial_j 
v^{(1)j}_\gamma +2 \frac{R}{1+R}\Psi^{(1)\prime} v^{(1)i}_\gamma
\nonumber \\ 
&-&2(\Phi^{(1)}+\Psi^{(1)}) \Phi^{(1),i} 
-\frac{1}{2(1+R)}(\Phi^{(1)}+\Psi^{(1)}) \Delta^{(1),i}_{00}
\nonumber \\
&-&\frac{R}{1+R} \partial^i v^{(1)2}_\gamma 
-\frac{3}{2}\frac{R}{1+R} \Delta^{(1)}_{00}\left(
\frac{{\cal H}}{1+R} v^{(1)i}_\gamma-\frac{1}{4} \frac{\Delta^{(1),i}_{00}}{1+R}
   \right)\, .
\end{eqnarray}
Here $\omega_i$ is the pure second-order metric perturbation of the $(0-i)$ metric tensor (which therefore does not appear in the first-order equation) while $\Pi^{(2)}_{ij}$ is 
the second-order quadrupole moment of the photons. If one takes the source term ${\cal S}^{i}_V$, Eq.~(\ref{SV}), in the squeezed limit and for $R=0$, one recovers exactly Eq.~(\ref{vtightsqueezed2nd})
once Eq.~(\ref{v2eqf}) is expressed in terms of the bolometric temperature $[\Delta^{(2)}_{00}-(\Delta^{(1)}_{00})^2]$. Notice that the second-order velocity continuity equation of the photon-baryon system in the squeezed limit is recovered only when $k_1 R \eta_{\rm rec} < 1$, i.e. when the timescale of the modulating long wavelength mode is much bigger than the typical timescale of the collision 
term in the tight coupling limit (in particular for $R\ll 1$). This does not come as a surprise,  the coordinate transformation we are adopting is not fully exact when $R$ is not much smaller than unity.

\section{The  bispectrum at recombination in the squeezed limit and the contamination to the primordial local NG}
Since we will be concerned with 
a signal-to-noise ratio $(S/N)$ dominated by   the maximum multipole a given 
experiment can reach, $\ell_{\rm max}\gg 1$, we can use the flat-sky approximation ~\cite{Hulensing} and write
for the bispectrum 

\be
\langle a(\vec{\ell}_1)a(\vec{\ell}_2)a(\vec{\ell}_3) \rangle 
   = (2\pi)^2\delta^{(2)}(\vec{\ell}_1+\vec{\ell}_2+\vec{\ell}_3) B(\ell_1,\ell_2,\ell_3)\, .
\ee 
Inspecting Eq. (\ref{bol}), the bispectrum gets two contributions. One from the intrinsically second-order term $\Theta^{(2)}$ and one from the term $(\Theta^{(1)})^2/2$. 

The first contribution can be calculated through the rescaling of the coordinates. 
We  make the simplifying assumption that  $\eta_{\rm rec}\gg \eta_{\rm eq}$ in such a way that the coordinate 
transformations (\ref{rescaling1}) and (\ref{rescaling2}) can be performed in a 
matter-dominated period, that is we take $\alpha=2$ and

\be
\label{rel}
{\bf x}\rightarrow e^{ -5\Phi_{\ell}/3}\,{\bf x}\, ,\,\,\,\, {\bf k}\rightarrow e^{5\Phi_{\ell}/3}\,{\bf k}\, ,\,\,\,\,
\eta_{\rm rec}\rightarrow e^{\Phi_{\ell}/3}\,\eta_{\rm rec}\, ,
\ee
is the transformation for modes which were outside the horizon at recombination, but are subhorizon at the time of observation. It is therefore enough to consider the rescaling acting on

\be
a(\vec{\ell})=\int\frac{{\rm d} {k}^z}{2\pi}\,e^{i {k}^z D_{\rm rec}}\,\Phi^{(1)}_{{\bf k}'}(0)\,\Delta_T(\ell,{k}^z)\, ,
\ee
where $\Delta_T(\ell,{k}^z)$ is the radiation transfer function and ${D}_{\rm rec} =(\eta_0-\eta_{\rm rec})$  is the distance to the surface of last scattering. We consider only the contribution
at recombination and therefore 

\begin{equation}
\label{aq}
        {\Delta}_T(\ell,k^z) = \frac{1}{D_{\rm rec}^2}  
          S(\sqrt{(k^z)^2 + \ell^2/ D_{\rm rec}^2},\eta_{\rm rec})\, ,\,\,S=\left(\frac{1}{4}\Delta^{(1)}_{00}+ \Phi^{(1)}+ {\bf v}^{(1)} \cdot {\bf n}\right)_{\eta=\eta_{\rm rec}}/\Phi^{(1)}_{{\bf k}}(0)\, . 
   \end{equation}
Here  ${k}^z$  is 
the momentum component
perpendicular  to the plane orthogonal to the line-of-sight and the superscript $^{'}$ reminds us that in the expression for ${k}$ we have to set $\vec{k}^{\parallel}=\vec{\ell}/{D}_*$, where $\vec{ k}^{\parallel}$ is the
component perpendicular to the line-of-sight.
As we have learnt  in the previous section,  the rescaling (\ref{rel}) reproduces the
second-order sources and the corresponding solutions starting from the first-order expression of the
temperature anisotropy. Therefore, we have to think of the expression (\ref{aq}) as a function of the rescaled coordinates in such a way that an expansion of the long mode will give the exact  second-order quantity
in the squeezed limit.   
Notice that  the rescaling (\ref{rel}) changes also the gravitational potential

\be
\Phi^{(1)}_{{\bf k}}\rightarrow e^{ -5\Phi_{\ell}}\,\Phi^{(1)}_{e^{5\Phi_{\ell}/3}{\bf k}}\, ,
\ee
and therefore $a(\vec{\ell})$ is not subject to any rescaling if one takes the Sachs-Wolfe large-scale limit in which the source $S$ reduces to $1/3$. 

Following Ref. \cite{BRRec}, we  mimic the effects of the transfer function on small scales as 

\be
{\Delta}_T(\ell,k^z)=\frac{a}{D_{\rm rec}^{2}}\,  e^{-1/2(\ell/\ell_*)^{1.2}}
e^{-1/2(|k_z|/ k_*)^{1.2}}\, ,
\ee 
{\it i.e.} a simple exponential and a normalization 
coefficent $a$ to be 
determined to match the amplitude of the angular power spectrum at the characteristic scale 
$\ell \simeq \ell_*=k_*D_{\rm rec}$.  As shown in Ref. \cite{BRRec}, the values  $\ell_*\simeq750$ and $a\simeq 3$ 
are able to account for the combined effects of ``radiation driving'', 
which occours at $\ell>\ell_{\rm eq}\simeq 160$ and boosts the angular power spectrum with respect to 
the Sachs-Wolfe plateau, 
and the effects of 
Silk damping which tend to suppress the CMB anisotropies for scales $\ell>\ell_{\rm D}\simeq 1300$. 
The combination of these 
effects produces a decrease in the angular power spectrum from a scale $\ell_*\simeq 750$.
The choice of the exponent $1.2$ derives from the study of the diffusion damping envelope in Ref. \cite{HuWhitedamping}.
The power spectrum in 
the flat-sky approximation is given 
by  

\be
\langle a(\vec{l}_1)a(\vec{l}_2) \rangle 
   = (2\pi)^2\delta^{(2)}(\vec{l}_{1}+\vec{\ell}_2) C(\ell_1)\, ,
   \ee
 with 
\be
C(\ell)=\frac{D_{\rm rec}^2}{(2 \pi)} \int {\rm d}k^z |{\Delta}_T(\ell,k^z)|^2\, P(k)\, .
\ee       
The exponential of the transfer function  allows to cut off the 
integral for $k\simeq k_*$ and one finds~\cite{BRRec}
\be
\label{Cl}
C(\ell)
\simeq a^2 \frac{A}{\pi}\frac{\ell_*}{\ell^3}\,
e^{-(\ell/\ell_*)^{1.2}}\, ,
\ee
which holds for $\ell\gg\ell_*$ and we have used  the amplitude of the  primordial gravitational potential 
power spectrum computed at first-order

\be
\langle \Phi^{(1)}({\bf k}_1) \Phi^{(1)}({\bf k}_2) \rangle = (2 \pi)^3 \delta^{(3)}
\big({\bf k}_1 + {\bf k}_2 \big) P(k_1)\, ,\,\,\,P(k)=A/k^3
\ee 
with amplitude $A=17.46 \times 10^{-9}$. 
To compute the bispectrum, we go to the squeezed limit
 $\ell_1\ll \ell_2,\ell_3$ (or $k_1\ll k_2,k_3$). In this case $\Phi^{(1)}_{{\bf k}_1}$ acts as a background
for the other two modes.  One can therefore compute the three-point function in a two-step process: first compute the 
two-point function in the background of $\Phi^{(1)}_{{\bf k}_1}$ and then the result from the correlation induced by the 
background field. Following Ref. \cite{Crec} (see also~\cite{CFKS}) and using the Sachs-Wolfe limit for the multipole $\ell_1$, this procedure leads to 

\be
\langle a(\vec{\ell}_2)a(\vec{\ell}_3)\rangle_{\Phi^{(1)}_{{\bf k}_1}}=\langle a(\vec{\ell}_2)a(\vec{\ell}_3)\rangle_0+5\,a(\vec{\ell}_2+\vec{\ell}_3)C(\ell_2)\frac{{\rm d}\ln \left[\ell_2^2C(\ell_2)\right]}{{\rm d}\ln \ell_2}\, .
\ee
The bispectrum from the intrinsically second-order term $\Theta^{(2)}$ therefore reads

\be
B_{\Theta^{(2)}}(\ell_1,\ell_2,\ell_3)=\Big<a(\vec{\ell}_1)\langle a(\vec{\ell}_2)a(\vec{\ell}_3)\rangle
\Big>=(2\pi)^2\,\delta^{(2)}(\vec{\ell}_1+\vec{\ell}_2+\vec{\ell}_3)\, 5\, \,C(\ell_1)C(\ell_2)\,  \frac{{\rm d}\ln \left[\ell_2^2C(\ell_2)\right]}{{\rm d}\ln \ell_2} \, .
\ee
Notice also  in all these computations one can safely neglect the rescaling of the conformal time at recombination $\eta_{\rm rec}$ up to terms $\eta_{\rm rec}/\eta_0\ll 1$.   

Next we compute the contribution from the term  $(\Theta^{(1)})^2/2$. In the squeezed limit it reads (see also Refs.~\cite{Nitta,review})

\be
B_{\frac{1}{2}\left(\Theta^{(1)}\right)^2}(\ell_1,\ell_2,\ell_3)=(2\pi)^2\,\delta^{(2)}(\vec{\ell}_1+\vec{\ell}_2+\vec{\ell}_3)\, 2\, \,C(\ell_1)C(\ell_2) \, .
\ee
The total bispectrum therefore is

\be
B_{\rm rec}(\ell_1,\ell_2,\ell_3)=(2\pi)^2\,\delta^{(2)}(\vec{\ell}_1+\vec{\ell}_2+\vec{\ell}_3) \,C(\ell_1)C(\ell_2)\left[2+5\,  \frac{{\rm d}\ln \left[\ell_2^2C(\ell_2)\right]}{{\rm d}\ln \ell_2}\right] \, .
\ee
Notice that if we consider the second-order CMB anisotropies on large-scales with all the modes on super-Hubble scales
at recombination (i.e. with all wavenumbers $(k_i \eta_{\rm rec} \ll 1)$, but inside the horizon today), then in the Sachs-Wolfe limit~\footnote{
This expression is valid only when all the wavelength modes are outside the horizon at recombination, and, e.g., when  written as a convolution in Fourier space,  it does not hold for an external wavenumber $k_1 \eta_{\rm rec} \ll 1$ with the two internal wavenumbers ${\bf k}'_2$ and ${\bf k}'_3$ inside the horizon.} 
\be
\label{Tsh}
\frac{\Delta T}{T}=\frac{1}{3}\Phi_{\rm rec}+\frac{1}{18}\left(\Phi^{(1)}_{\rm rec} \right)^2\, . 
\ee
The term $({\Phi^{(1)2}_{\rm rec}}/18)$ corresponds in this limit exactly to the $(\Theta^{(1)2}/2)$ term in Eq.~(\ref{bol}). The result~(\ref{Tsh}) was first obtained  in~Ref.~\cite{GI} (see also the discussion that followed in Ref.~\cite{B}). This additional contribution to primordial non-Gaussianity from 
the second-order gravitational potential in $(\Phi_{\rm rec}=\Phi^{(1)}_{\rm rec}+\Phi^{(2)}_{\rm rec}/2)$ and the term $({\Phi^{(1)2}_{\rm rec}}/18)$ (see Refs.~\cite{GI,beyond}) can contribute to a contamination of the primordial local non-Gaussianities only on the very largest scales,  $2\lsim \ell \lsim 200$. Therefore they actually lead to a small (negligible) contribution 
to the total level of contamination (see the quantity $f^{\rm con}_{\rm NL}$ defined below). 

Our goal now is to estimate the level of 
degradation that the NG from recombination in the squeezed limit  causes
on the possible measurement of the local  primordial bispectrum. 
A rigorous procedure is to define the Fisher matrix (in flat-sky approximation) as 
\be 
F_{ij}=\int d^2 \ell_1 d^2 \ell_2  d^2 \ell_3 
\,\delta^{(2)}(\vec{\ell}_{1}+\vec{\ell}_{2}+\vec{\ell}_{3})\,\frac{
B_{i}(\ell_1,\ell_2,\ell_3)\, B_{j}(\ell_1,\ell_2,\ell_3)}{6\, C(\ell_1)\,C(\ell_2)\, C(\ell_3)}\, ,
\ee
where $i$ (or $j$)$=({\rm rec},{\rm loc})$, and to define the signal-to-noise ratio for a component $i$, 
$(S/N)_i=1/\sqrt{F^{-1}_{ii}}$,  and the degradation parameter $d_i=F_{ii} F^{-1}_{ii}$, due to the correlation 
bewteen the 
different components $r_{ij}=F^{-1}_{ij}/\sqrt{F^{-1}_{ii}F^{-1}_{jj}}$. In order to measure the contamination to the primordial bispectra 
one can define that effective non-linearity parameter $f^{\rm con}_{\rm NL}$ which minimizes the $\chi^2$ defined as 
\begin{eqnarray}
\chi^2=\int d^2 \ell_1 d^2 \ell_2  d^2 \ell_3 
\,\delta^{(2)}(\vec{\ell}_{1}+\vec{\ell}_{2}+\vec{\ell}_{3})\,
\frac{\left[f^{\rm con}_{\rm NL} B_{\rm loc}(\ell_1,\ell_2,\ell_3;f_{\rm NL}^{\rm loc}=1 )
- B_{\rm rec}(\ell_1,\ell_2,\ell_3)\right]^2}{6\, C(\ell_1)\,C(\ell_2)\, C(\ell_3)}\, , \nonumber \\
\end{eqnarray}
to find 
\begin{equation}
\label{effr}
f^{\rm con}_{\rm NL}= \left.\frac{F_{\rm rec,loc}}{F_{\rm loc ,loc}} \right|_{f_{\rm NL}^{\rm loc}=1}\, .
\end{equation}
In multipole space the bispectrum induced by a local primordial NG in the squeezed limit is given by 

\be
\label{eq:loc}
B_{\rm loc}(\ell_1,\ell_2,\ell_3) = -6\,f_{\rm NL}^{\rm loc}   \left[C(\ell_1)C(\ell_2)+{\rm cycl.}\right]
 \, .
\ee
Performing the integrals in the multipoles and using Eq. (\ref{Cl}), we find
\be
\label{tot1}
f^{\rm con}_{\rm NL}\simeq -\frac{1}{6}+\frac{5}{12}\left[1+0.75 \left(\frac{\ell_{\rm max}}{\ell_*}\right)^{1.2}-\left(\frac{\ell_{\rm min}}{\ell_{\rm max}}\right)^2 \left(1+0.75 \left(\frac{\ell_{\rm min}}{\ell_{*}} \right)^{1.2} \right) \right]\simeq 0.91\, ,
\ee
where we have taken $\ell_{\rm max}=2000$, $\ell_*=750$ and $\ell_{\rm min}= 1200$, see Ref.~\cite{BRRec}. 

Let us also remark that in Refs.~\cite{BRRec,review} the non-linear evolution on small scales (for all the wavenumbers ($k_i \eta_{\rm rec} \gg1$) at recombination) 
of the second-order gravitational potential  also produces a small contamination to the local 
non-Gaussianity of the order of $f^{\rm con}_{\rm NL}= 0.3$. This contribution should be added to Eq.~(\ref{tot1}).

\section{Conclusions and remarks} 
In this paper we have analytically estimated the level of non-Gaussianity produced by the non-linear evolution of the photon-baryon system at recombination in the squeezed limit.  While a
contamination ${\mathcal O}(5)$ was numerically obtained in~\cite{frII}, we do find that the total contamination to the primordial local non-Gaussianity is smaller and not within the reach of present experiments.
Our main goal was to provide a clear and simple way to understand the physical origin of such a contamination. We have reached the goal using a  simple rescaling of the local coordinates of a 
perturbed FRW universe, which explains in a very transparent way the large-scale modulation of the perturbations which is generally 
at the origin of a squeezed non-Gaussian signal. 
Let us make some brief remarks about the validity of our results. It is clear that they are valid up to terms $\mathcal{O}(\Phi^{(1)}_{{\bf k}_2} \nabla \Phi^{(1)}_{{\bf k}_1})$, 
which vanish in the exact squeezed limit, $k_1 \rightarrow 0$. Of course such contributions will arise for the perturbation modes ${\bf k}_1$ that were not much outside the Hubble radius 
at recombination, e.g. for $k_1\simeq k_{\rm eq}$. For example the initial condition for the second-order gravitational potential would get a small correction (see Eq.~(269) of~\cite{review}) 
from the initial condition (with vanishing primordial non-Gaussianity)
\begin{equation}
\label{initsqueezedfull}
\Psi^{(2)}_{\bf k}(0)=11 \frac{F({\bf k}_1, {\bf k}_2, {\bf k})}{k^2}  \Psi^{(1)}_{{\bf k}_1}(0)\Psi^{(1)}_{{\bf k}_2}(0)\, , 
\end{equation}
where the function $F({\bf k}_1, {\bf k}_2, {\bf k})$ would vanish in the exact squeezed limit (see its expression in Eq.~(242) of~\cite{review}. It is easy to check that, if for example, one takes 
$k_1=k_{\rm eq}$ and $k_2=15 k_{\rm eq}$ (the latter corresponding to $\ell_2=2000$) one finds that the contamination to the primordial local NG induced from the correction of the $F$ term 
in Eq.~(\ref{initsqueezedfull}) is $|f^{\rm con}_{\rm NL}({\rm loc})| \simeq 0.1$. Indeed this must also be intended as an upper limit, since this correction does not fully correlate with the local type 
$f_{\rm NL}^{\rm loc}$.  This example is to show that the corrections coming from the perturbation modes  $k_1=k_{\rm eq}$ is actually subdominant. 

\section*{Acknowledgments}
This research has been partially supported by the ASI/INAF Agreement I/072/09/0 for the Planck LFI Activity of Phase E2.

\section*{Note added}
When completing this work, we have become aware of a similar work by P. 
Creminelli, C. Pitrou and F. Vernizzi. Our results, when overlap is 
possible, agree with theirs. We thank them for useful correspondence.


\end{document}